\documentclass[manuscript]{aastex}

\shorttitle{LSST yield of eclipsing binaries}
\shortauthors{Pr\v sa, Pepper and Stassun}

\defcitealias{LSSTbook}{LSST Science Book}

\begin{document}

\title{Expected Large Synoptic Survey Telescope (LSST) Yield of Eclipsing Binary Stars}

\author{Andrej Pr\v sa}
\affil{Villanova University, Department of Astronomy and Astrophysics, 800 E Lancaster Ave, Villanova PA 19085}
\email{andrej.prsa@villanova.edu}

\and

\author{Joshua Pepper}
\affil{Vanderbilt University, Physics and Astronomy Department, 2201 West End Ave, Nashville TN 37235}

\and

\author{Keivan~G.~Stassun}
\affil{Vanderbilt University, Physics and Astronomy Department, 2201 West End Ave, Nashville TN 37235}
\affil{Fisk University, Department of Physics, 1000 17th Ave. N., Nashville, TN 37208}


\begin{abstract}
In this paper we estimate the Large Synoptic Survey Telescope (LSST) yield of eclipsing binary stars, which will survey $\sim$20,000 square degrees of the southern sky during the period of 10 years in 6 photometric passbands to $r\sim 24.5$. We generate a set of 10,000 eclipsing binary light curves sampled to the LSST time cadence across the whole sky, with added noise as a function of apparent magnitude. This set is passed to the Analysis of Variance (AoV) period finder to assess the recoverability rate for the periods, and the successfully phased light curves are passed to the artificial intelligence-based pipeline {\tt ebai} to assess the recoverability rate in terms of the eclipsing binaries' physical and geometric parameters. We find that, out of $\sim$24 million eclipsing binaries observed by LSST with $S/N>10$ in mission life-time, $\sim$28\% or 6.7 million can be fully characterized by the pipeline. Of those, $\sim$25\% or 1.7 million will be double-lined binaries, a true treasure trove for stellar astrophysics.
\end{abstract}



\section{Introduction}

The importance of eclipsing binary stars (EBs) can hardly be overstated. Their analysis provides calibration-free physical properties of stars (i.e. masses, radii, surface temperatures, luminosities; cf.~\citealt{torres2010}); precise stellar distances \citep{guinan1998}; stellar ages \citep{thompson2010}; and stringent tests of stellar evolution models \citep{stassun2009}. The products of state-of-the-art EB modeling are seminal to many areas of astrophysics, including: calibrating the cosmic distance scale; mapping clusters and other stellar populations (e.g. star-forming regions, streams, tidal tails, etc) in the Milky Way; determining initial mass functions and studying stellar population theory; understanding stellar energy transfer mechanisms (including activity) as a function of temperature, metallicity and evolutionary stage; calibrating stellar color-temperature transformations, mass-radius-luminosity relationships, and other relations basic to a broad array of stellar astrophysics; studying stellar dynamics, tidal interactions, mass transfer, accretion, chromospheric activity, etc.

The upcoming decade promises to yield astronomical data that are unprecedented both in quantity as well as in accuracy. The Panoramic Survey Telescope \& Rapid Response System \citep[Pan-STARRS;][]{kaiser2002} is conducting their first pilot study \citep[][PS1]{jedicke2007} in which a quarter of all Pan-STARRS targets are observed in an effort to test the telescope, camera and processing software. Although the main objective of the upcoming PS2 stage of Pan-STARRS is to detect potentially hazardous objects in the Solar system, the by-product will be a 6-band photometric database of $\sim$20 billion stars, each observed 20-60 times, visible from Hawaii \citep{kaiser2005}. It is scheduled to see first light in the Fall of 2012. The survey has no restrictions on operation duration. Gaia \citep{lindegren2008} is the next ESA cornerstone mission scheduled to launch in 2012; it will observe $\sim$1 billion stars, each about 70 times during its 5-year mission. While Gaia is primarily an astrometric mission, the observations will be complemented by low-resolution spectrophotometric data in the 330-1000\,nm wavelength range, the $G$ passband photometry from the unfiltered light in the astrometric field and, for the stars brighter than 17-mag, radial velocity measurements \citep{jordi2010}. The expected yield of eclipsing binary stars has been studied by \citet{munari2001} who estimate that Gaia will observe $\sim$100,000 double-lined eclipsing binaries. Already, {\it Kepler} \citep{borucki2010} has detected and characterized 1879 eclipsing and ellipsoidal binaries in a sample of $\sim 156,000$ stars \citep{prsa2011}.

The Large Synoptic Survey Telescope \citep[][LSST]{ivezic2008} is a large, wide-field system designed to obtain multiple images covering the southern sky. LSST is scheduled to see first light in mid-2018. The current baseline design, with an effective 6.5\,m primary mirror, a 9.6\,deg${}^2$ field of view, and a 3.2 Gigapixel camera, will allow about 20,000 square degrees of sky to be covered using pairs of 15-second exposures in two out of six SDSS ($ugrizy$) photometric bands every three nights on average, with typical 5$\sigma$ depth for point sources of $r\sim24.5$. The \citetalias{LSSTbook} (\S6.10, pg.~171) includes a discussion on the importance of EBs to the LSST science case, and includes a preliminary estimate of the EB yield. It proposes a metric based on the number of visits (pairs of 15-s observations) in the eclipses and provides qualitative results that indicate LSST's near-perfect efficiency for short-period binaries, $\sim$50\% efficiency for periods up to 30 days, and a $\sim$10\% efficiency for the longer periods. Here we revise the estimate in significantly greater detail and provide quantitative results.

The paper structure is as follows: in \S2 we present the details of the simulation, in \S3 we discuss the recoverability rate for the ephemerides and for physical parameters; in \S4 we summarize the results and conclude in \S5.

\section{Model LSST Light Curves and Analysis}

In this section we describe our procedures to estimate LSST's EB yield. First we created a set of model EB light curves employing PHOEBE \citep{prsa2005}, a \citet{wd1971} based eclipsing binary modeling suite. Next, we sampled these light curves according to LSST's so-called universal cadence \citep{cook2009} over a nominal 10-year life span of the mission. We then added noise as a function of apparent magnitude, where the distribution of magnitudes was drawn from the Sloan Digital Sky Survey (SDSS) luminosity function. The light curves were then passed to a standard period finder to determine the ephemerides, and phased light curves were processed by our neural network-based engine, {\tt ebai}, to automatically estimate principal parameters of the systems. Finally, we compared these estimated parameters with the ones actually used to create the sample, which provided us with the LSST efficiency metric for the EB yield. We now describe these steps in turn.

\subsection{Input Light Curve Parameters}

We synthesized 10,000 detached eclipsing binary light curves, described by their ephemerides (HJD${}_0$ and period $P_0$) and 5 principal parameters: $T_2/T_1$, $\rho_1+\rho_2$, $e \sin \omega$, $e \cos \omega$ and $\sin i$. The temperature ratio $T_2/T_1$ serves as a proxy to the surface brightness ratio $B_2/B_1$ that directly determines the ratio of depths of both eclipses. The sum of fractional radii $\rho_1+\rho_2 \equiv (R_1+R_2)/a$ determines the baseline width of the eclipses. The radial eccentricity projection $e \sin \omega$, where $\omega$ is the argument of periastron, determines the ratio between the secondary and the primary eclipse widths. The phase separation of the eclipses is proportional to the tangential eccentricity projection $e \cos \omega$, and the overall amplitude of the light curve, as well as the shape of eclipses ({\sf U}- or {\sf V}-shaped), are determined by $\sin i$. The limb darkening parameters were automatically interpolated from Pr\v sa et al.~(2011; in prep.) based on component temperature and effective gravity, and gravity darkening exponents to 1.0 and 0.32 for radiative and convective envelopes, respectively. For a thorough discussion on the choice of principal parameters refer to \citet{prsa2008}.

\begin{figure}
\begin{center}
\includegraphics[width=0.8\textwidth]{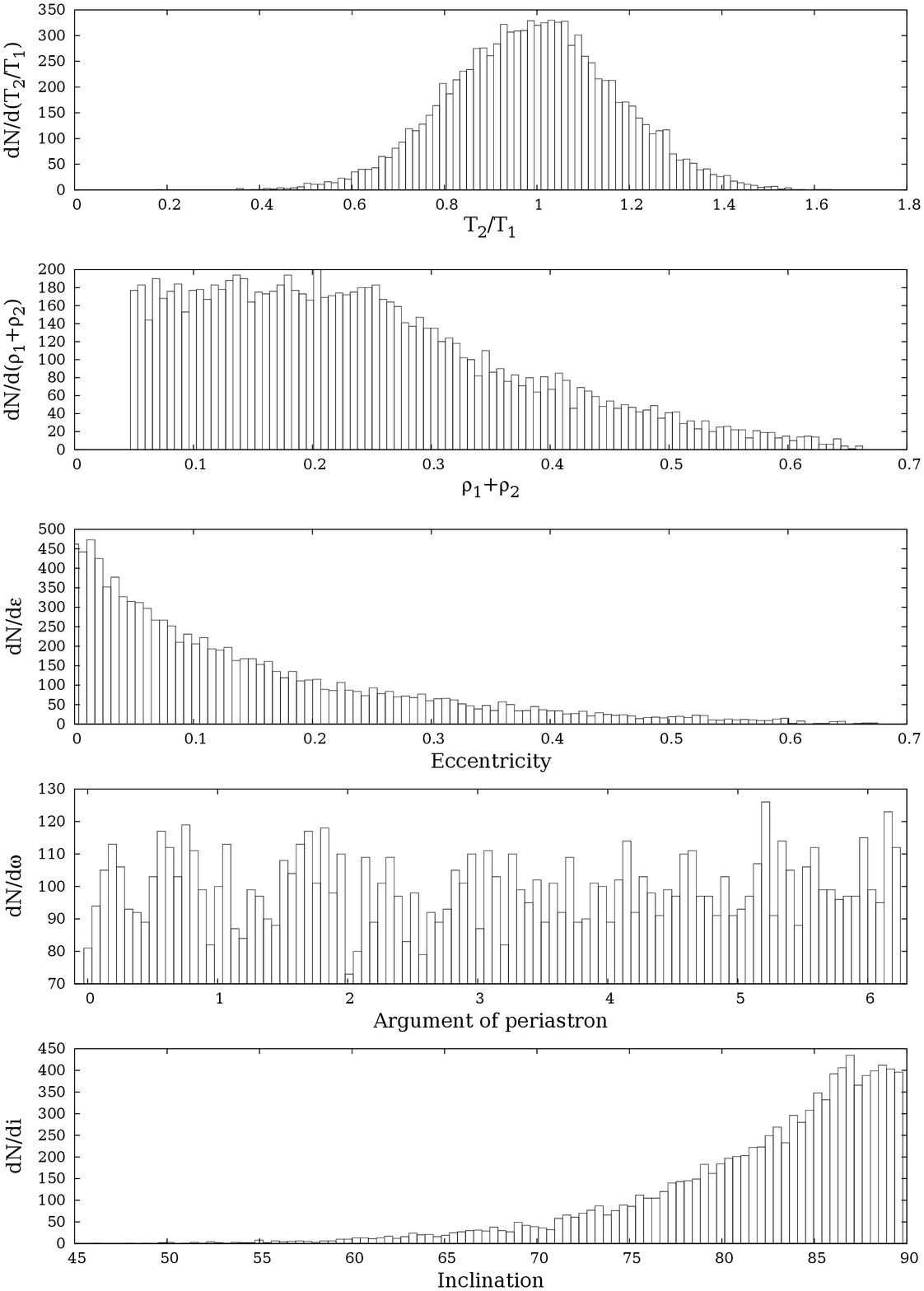} \\
\caption{Theoretical probability distribution functions for $T_2/T_1$, $\rho_1+\rho_2$, $e$, $\omega$, and $i$. Principal parameters of the training set and of the test set for 10,000 EBs were sampled from these distributions. \label{fig:dists}}
\end{center}
\end{figure}

The values of principal parameters were randomly sampled from the following probability distribution functions (PDFs; cf.~Fig.~\ref{fig:dists}): 

\begin{itemize}
\item $T_2/T_1$ is sampled from a normal distribution $\mathcal G(1.0, 0.18)$;
\item $P_0$ is sampled from a log-uniform distribution $[-1,4]$;
\item HJD${}_0$ is sampled from a uniform distribution $[0,P_0]$;
\item $\rho_1+\rho_2$ is sampled from a uniform distribution $[0.05, \delta_\mathrm{max}-0.05]$, where $\delta_\mathrm{max}$ is the morphology constraint parameter that depends exponentially on the value of $\log P$:
$$ \delta_\mathrm{max} (\log P_0) = 0.7 \exp \left( -\frac{1+\log P_0}{4} \right); $$
\item eccentricity $e$ is sampled from an exponential distribution $\mathcal E(0.0, \epsilon_\mathrm{max})$, where $\epsilon_\mathrm{max}$ is the attenuation parameter that depends exponentially on the value of $\rho_1+\rho_2$:
$$ \epsilon_\mathrm{max} (\rho_1+\rho_2) = 0.35 \exp \left( -\frac{\rho_1+\rho_2-0.05}{1/6} \right); $$
\item argument of periastron $\omega$ is sampled from a uniform distribution $[0, 2\pi]$;
\item sine of the inclination $\sin i$ is sampled from a uniform distribution $[\sin i_\mathrm{grazing}, 1]$, where $i_\mathrm{grazing}$ is the inclination of a grazing eclipse:
$$
\sin i_\mathrm{grazing} = \sqrt{1-(\rho_1+\rho_2)^2}.
$$
\end{itemize}

The reason why we did not simply use uniform distributions in all parameters and thus cover the whole parameter space is to avoid unphysical light curves. Such light curves, when they comprise a significant fraction of the whole sample, have a notable adverse impact on the network's recognition ability, especially for second-order parameters such as $e \sin \omega$ and $e \cos \omega$ \citep{prsa2008}.

\subsection{Input Light Curve Cadence}

LSST will survey the sky according to the so-called \emph{universal} cadence. This scanning model is optimized for the uniform observed depth of $r \sim 24.5$ and a uniform number of visits across 20,000 square degrees of the southern sky \citepalias[\S 2.1]{LSSTbook}. A \emph{visit} is defined as a pair of 15\,s exposures, performed back-to-back, with a 4\,s readout separation, to aid in cosmic ray rejection. Most fields will be observed twice during the nightly run, with visits separated by 15-60 minutes. This is done to maximize the sensitivity to motion of solar system objects but also to benefit the detection of short period stellar variability. The universal cadence excludes visits of $\sim$1,000 square degrees around the galactic center because of crowding.

To simulate this universal cadence, we made use of the \emph{Operations Simulator} {\tt opsim 1.29} that serves to evaluate the suitability of the scanning model and to quantify the yield for individual scientific goals of the survey \citepalias[\S3.1]{LSSTbook}. The simulator provides an array of heliocentric Julian dates (HJD) at which a given field will be observed.

We split the southern sky uniformly in declination:

$$ \delta_i = \frac{i-0.5}{N} \times 100^\circ - 90^\circ \qquad \textrm{for } i = 1 \dots N, $$
where $N$ is the number of declination bands. The $i$-th band is then split into $M_i$ right ascensions:

$$ M_i = 1 + \mathrm{int}\,(1.3 \times 2N \sin \delta_i); $$
$$ \alpha_j = \frac{j-0.5}{M_i} \times 360^\circ \quad \textrm{for } j = 1 \dots M_i. $$

\begin{center}
\begin{figure}
\includegraphics[height=\columnwidth, angle=270]{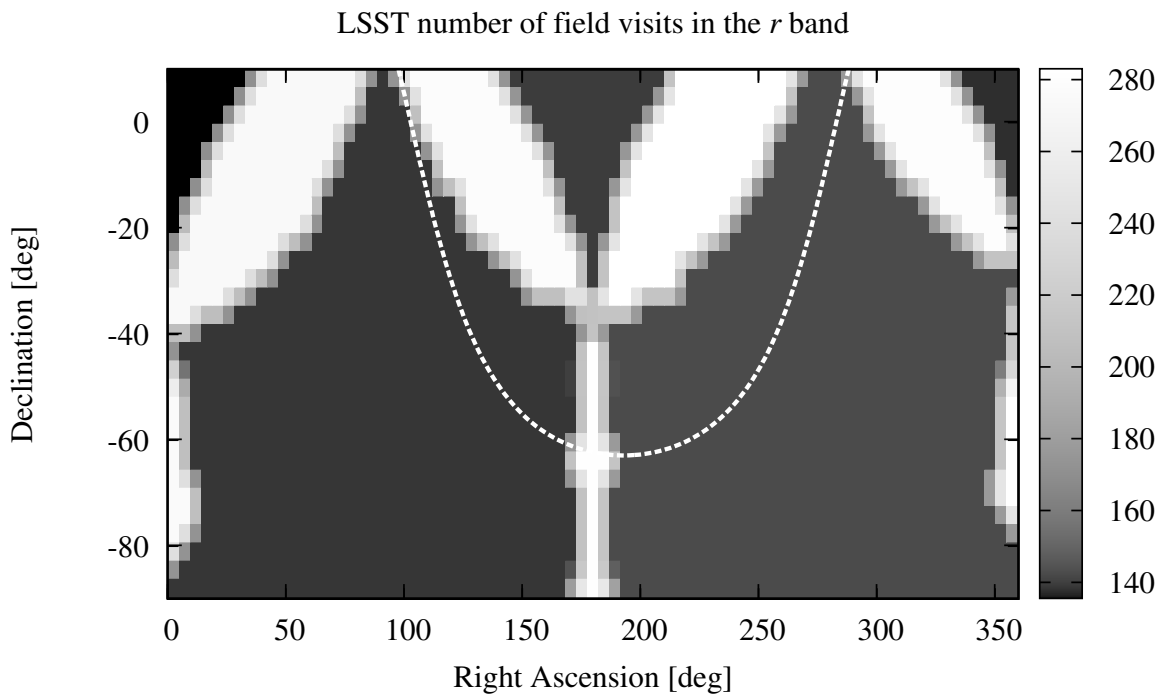} \\
\caption{The number of LSST $r$-band visits (pairs of 15-s exposures) across the sky, according to the {\tt opsim1.29} universal cadence. The operations simulator has a sophisticated model of the telescope and dome and outputs the cadences (heliocentric Julian dates) per passband as a function of right ascension and declination of the field. The nominal sky coverage is 20,000 square degrees across the southern hemisphere, and averages to $\sim$230 visits per field. The dashed line depicts the galactic equator. \label{EB_fig_skymap}}
\end{figure}
\end{center}

\begin{center}
\begin{figure}
\includegraphics[height=\columnwidth, angle=270]{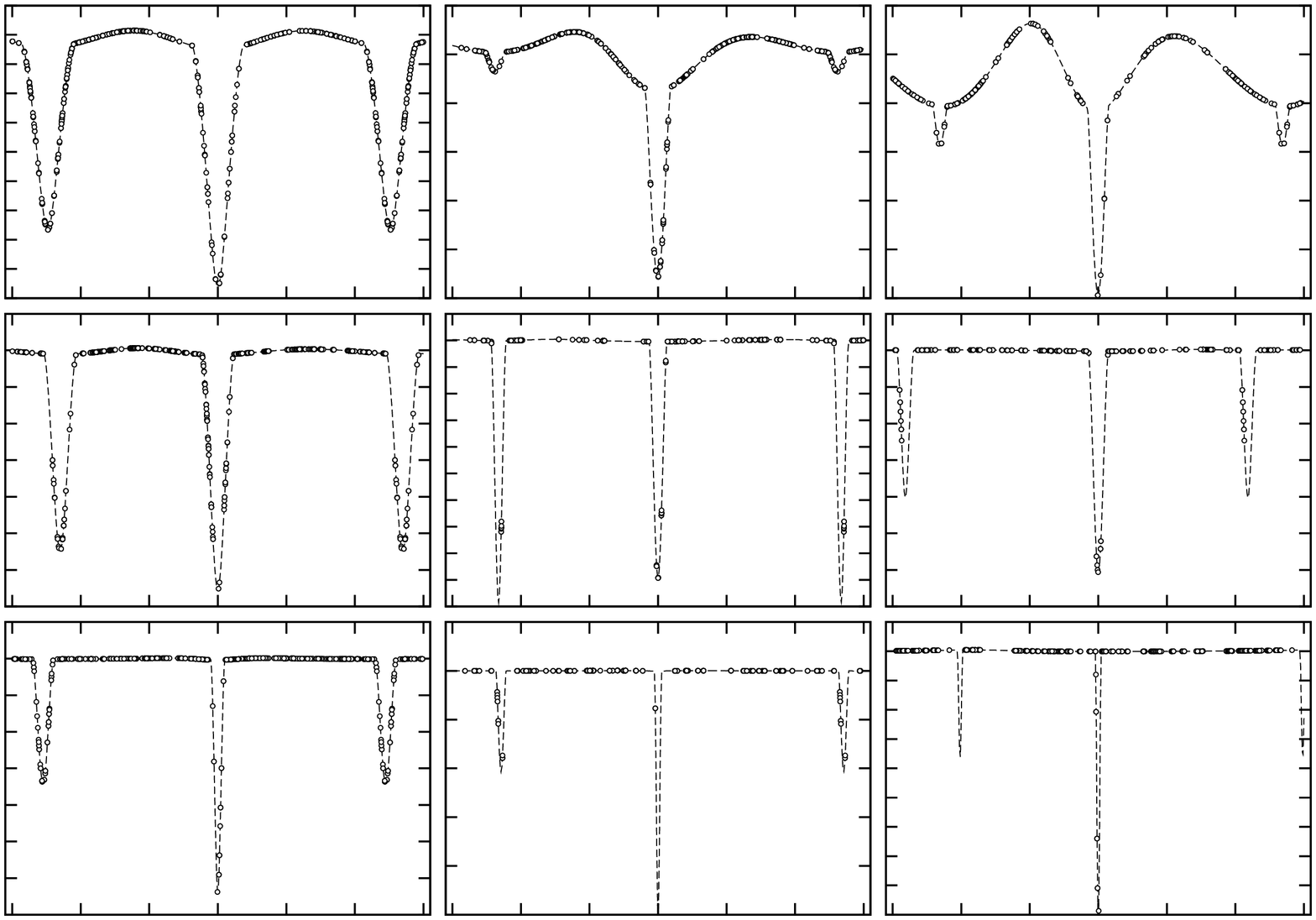} \\
\caption{Nine eclipsing binary light curves depicting the LSST cadence. The dashed line is a theoretical curve with no noise or period uncertainty added. The points are individual visits in the $r$ band. The panels are arranged from the shortest (top left) to the longest period (bottom right): 1.42-d, 1.80-d, 2.06-d, 2.53-d, 6.31-d, 6.72-d, 15.2-d, 57.9-d, and 109-d. The phase coverage becomes a problem towards longer periods: there is a single data point acquired during the primary eclipse of the 57.9-d binary, and no data points during the secondary eclipse of the 109-d binary. For detached EBs it is the number of data points in eclipse that predominantly govern period and parameter determinability. \label{fig:lightcurves}}
\end{figure}
\end{center}

The factor $1.3$ in the expression for $M_i$ makes the distribution in $\alpha$ 30\% denser to account for the pronounced universal cadence variability along $\alpha$. Sky partitioning in this way yielded 1558 fields for $N=30$, covering all right ascensions and declinations between $-90^\circ$ and $10^\circ$. Fig.~\ref{EB_fig_skymap} depicts the number of visits per field in the $r$ band.  The numbers typically vary from almost 600 points per lightcurve close to the celestial equator, down to 300 points per lightcurve at higher declinations. Fig.~\ref{fig:lightcurves} shows several examples of the LSST phase coverage on EB light curves with different orbital periods.

\subsection{Luminosity Function and Noise Model}

After the application of the LSST cadence to the model lightcurves, we then apply a combination of apparent luminosity function and noise model to obtain data similar to what we expect to emerge from the LSST data pipeline.  We first derive a luminosity function from the SDSS, taking a random selection of stars from the SDSS Data Release 7 (DR7) \citep{sdssdr7}.  We randomly selected 18,000 stars at all magnitudes and sky positions from the DR7 footprint.  We then fit a simple polynomial to the distribution of $r$-band magnitudes to model the luminosity function of field stars, and randomly assign apparent magnitudes to our sample lightcurves with a distribution based on the SDSS luminosity function in the range $16 <  r < 22$.  

We then apply Gaussian white noise to the lightcurves.  To match the assigned magnitude to an expected photometric error, we use a theoretical LSST noise model (S. Howell 2009, private communication).  That noise model is shown in Table \ref{tab_noidemodel}.

\begin{table}[t]
\begin{center}
\caption{Theoretical LSST Noise Model (Howell 2009, priv.~comm.)}
\begin{tabular}{cc}
$r$ magnitude & Expected per-point \\
 & RMS error (magnitude)$^{a}$ \\
\hline
16.0000 &  0.0011 \\
16.5000 &  0.0014 \\
17.0000 &  0.0018 \\
17.5000 &  0.0023 \\
18.0000 &  0.0030 \\
18.5000 &  0.0039 \\
19.0000 &  0.0051 \\
19.5000 &  0.0070 \\
20.0000 &  0.0097 \\
20.5000 &  0.0138 \\
21.0000 &  0.0202 \\
21.5000 &  0.0303 \\
22.0000 &  0.0461 \\
\hline
\multicolumn{2}{l}{${}^a$ Based on an exposure time of 15s.} \\
\end{tabular}
\label{tab_noidemodel}
\end{center}
\end{table}

\section{Recovery of EB Parameters from LSST Lightcurves}

With a set of EB lightcurves representative of what LSST will observe, our next step is to determine how well we can recover the physical parameters of the EBs.  We attempt this recovery with two separate tests.  First, we apply an off-the-shelf period-finding algorithm to obtain the correct period and phase the lightcurves.  Second, we take the lightcurves for which the true period was recovered within a specified tolerance, and use a neural network-based method to attempt to recover the physical parameters of the system. Note that we do not attempt to simulate the performance of light curve classifiers; rather, we assume that automated light curve classification is done successfully (see \citealt{debosscher2007} for general efficiency estimates). We plan to explore the classification success in detail in the follow-up paper.

\subsection{Period Recovery} \label{sec:per_recover}

Successful determination of the orbital period depends on many factors, and a variety of different methods have been utilized to identify their periods, such as analysis of variance (AoV; \citealt{s-c89}) or box-fitting least-square (BLS; \citealt{Kovacs2002}) (see \citealt{devor2005,hartman2008,hartman2009} for examples).  For the purposes of this analysis, we use the AoV method only.  We acknowledge that other methods will likely be able to improve upon the detection efficiencies we find here, especially in different regions of parameter space.  We therefore present the following results as a lower limit in terms of period recovery fraction.

In our implementation, we utilize the {\tt vartools}\footnote{http://www.cfa.harvard.edu/~jhartman/vartools/} utility \citep{hartman2009}.  We use the AoV algorithm to search for periodicities in all 10,000 lightcurves, with a period search range of 0.5 to 1000 days, and \emph{subsample} and \emph{finetune} parameters of 0.1 and 0.01, respectively.  We record the output periodigram of each lightcurve.

The criterion we use to determine a ``successful'' period recovery is the relation $k = \delta\tau/P^2$ (Oluseyi 2010, private communication).  This formula states that for any points in a phase-folded lightcurve, the maximum offset from their correct phase $k$, expressed as a fraction of the true period, is the fractional offset between the measured period $P_m$ and the true period $P$, times the number of cycles in the lightcurve.  The fractional offset is $f=\delta/P$, where $\delta = |P_m - P|$, and the number of cycles is the lightcurve duration $\tau$ divided by the true period $P$.  The parameter $k$ reflects the maximum offset within the phased lightcurve of the most discrepant points.  By testing various values for $k$, we find that a requirement of $k<10\%$ returns the largest number of correctly recovered periods.  We thus require that $k<10\%$, so $\delta < 0.1 P^2/\tau$.  To determine whether an EB is recovered via the AoV method, we consider the period of the highest peak in the periodigram ($P_k$).  Since EB lightcurves typically show more power at half the true period than the true period itself, we use twice the peak period as the matching value.  Thus, we define $P_m = 2P_k$.

\begin{figure}[ht!]
\begin{center}
\includegraphics[height=\columnwidth, angle=0]{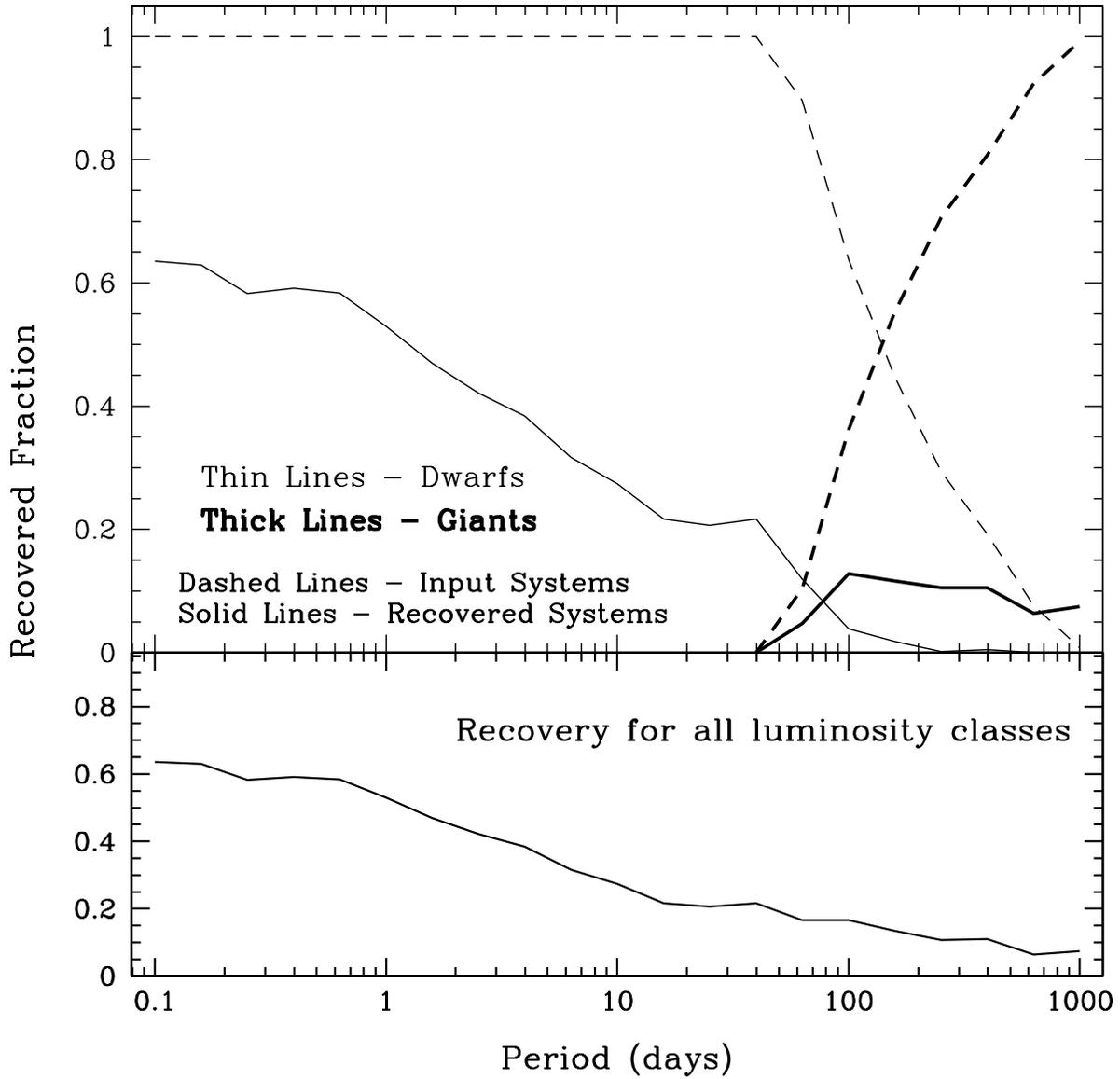} \\
\end{center}
\caption{The period recovery rate of eclipsing binary stars with the AoV technique, based on the luminosity class of the components and orbital period.  The period recovery criterion is $k<10\%$.  See \S \ref{sec:per_recover} for further details of the recovery criteria.}
\label{fig:aov1}
\end{figure}

We find that successful period recoverability is strongest for short period EBs and declines relatively evenly in $\log P$. This is expected since phase coverage for short period EBs is more complete than for long period EBs (cf.~Fig.~\ref{fig:lightcurves}). The results from the AoV analysis for different EB types are shown in Figure \ref{fig:aov1}. In this plot, we split up the input lightcurve sample according to the physical parameters of the EBs, based on the parameter combination ($\rho_1 + \rho_2) P^{2/3}$, which is roughly the sum of the radii of the components, $\approx (R_1 + R_2)$.  We separate the sample into Dwarfs ($R_1 + R_2 \leq 5 R_{\odot}$) and Giants ($R_1 + R_2 > 5 R_{\odot}$).  The dashed lines show the distribution of input systems and the solid lines show the recoverability, as a function of period.

Any attempt at period recovery for ground-based data typically encounters aliasing effects at integral fractions and multiples of 1 day.  For systems where the true period is very close to those periods, recovery tends to decrease strongly.  In Fig \ref{fig:aov1}, the resolution in period space is too large to see significant aliasing effects. While such effects are inherently present in LSST data, they are actually marginal due to the LSST cadence. Because of the combination of infrequent sampling (1-3 visits/night every $\sim$20 nights) and the very long time baseline (10 years), aliasing effects at the diurnal cycle are fairly small compared to higher-cadence and/or single-season campaigns.

\begin{figure}[ht!]
\begin{center}
\includegraphics[width=0.8\textwidth]{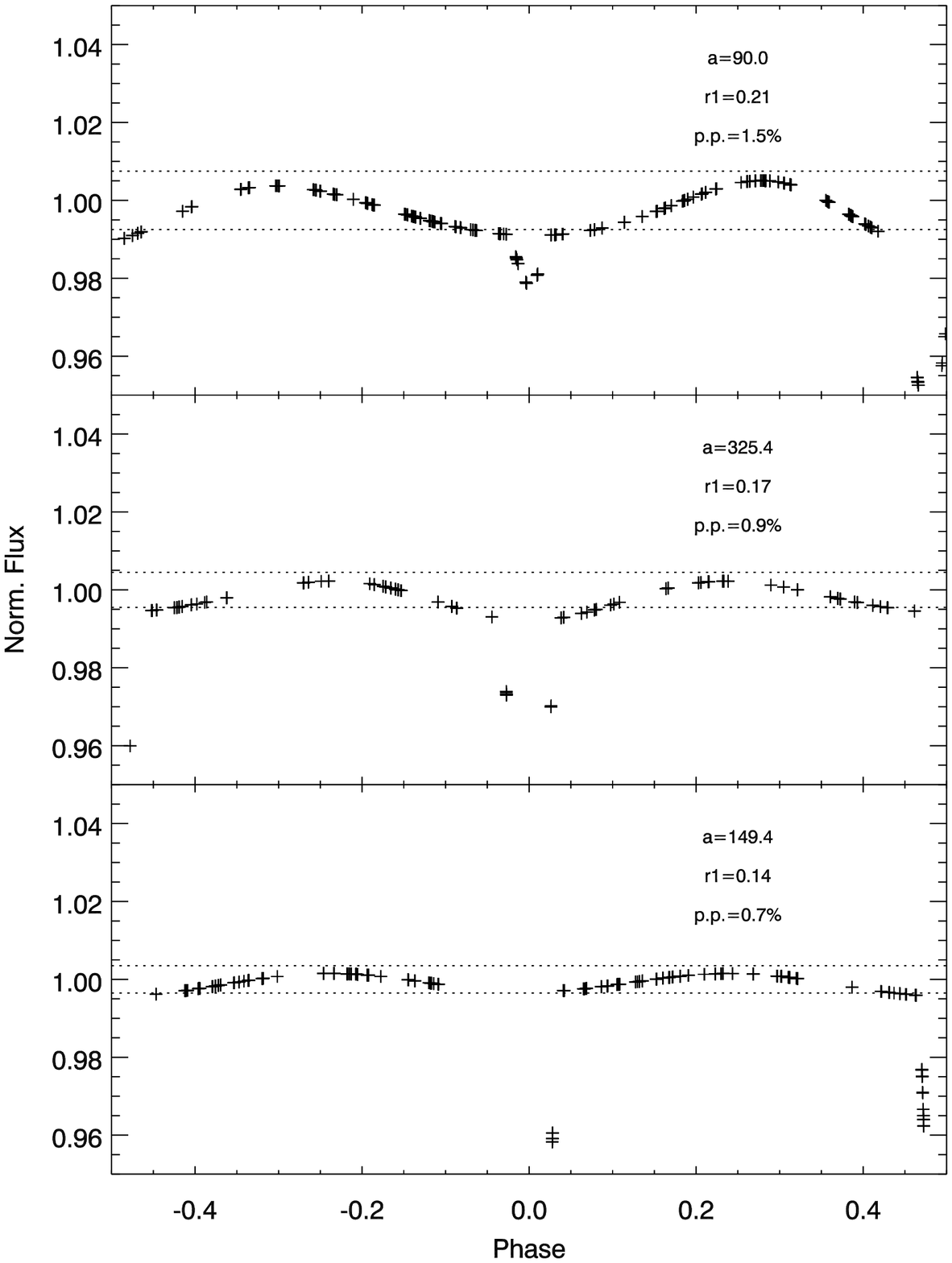} \\
\end{center}
\caption{
\label{fig:ellipvar}
Simulated noiseless light curves for three EBs with evolved (i.e., red giant) primary stars that present ellipsoidal variations despite having large semi-major axes (i.e., long orbital periods). In each panel is listed the EB semi-major axis (in solar radius units), the relative radius of the primary star ($R_1/a$), and the predicted peak-to-peak variations due to ellipsoidal effects. Horizontal dotted lines show the predicted ellipsoidal amplitude also. The vertical scale has been set to emphasize the low-level ellipsoidal variation; some data points in eclipse are lost below the figure in each case.
}
\end{figure}

It may seem surprising that we are able to accurately recover EB periods as long as $\sim$1000~d. We find that in many of these cases, particularly where one of the stellar components is a giant, it is the ellipsoidal variations that provide the strong observational handle on the system's period. In an EB where one component is a giant star, the giant can fill a large fraction of the semi-major axis even though the orbital period may be quite long. The proximity of the companion star can then raise tidal bulges on the giant star that lead to small but observable light curve variations on the orbital period. More quantitatively, the expected peak-to-peak ellipsoidal variation is $\approx 2 M_2/M_1 (R_1 \sin i/a)^3$ \citep{vankerkwijk2010}.  Figure~\ref{fig:ellipvar} shows three example light curves from our simulations that illustrate the above effects. In each panel, we list the semi-major axis (in solar radius units), the relative size of the primary star ($R1/a$), and the level of expected peak-to-peak variations.  For example, the middle panel illustrates the case of an EB with $a=325.4$~R$_\odot$ ($P_{\rm orb}=334$~d) and $R_1=55.3$~R$_\odot$. The expected level of ellipsoidal variation is $\approx$1\%, and this is indeed what we observe in the light curve. For the minimum photometric signal-to-noise adopted in our simulations, variability amplitudes of $\sim$1\% or larger should yield easily recovered periods by LSST.

We note that the relative proportion of giants in our test lightcurves (20\%) is slightly higher than the proportion in the Galaxy (5\% - 15\%, see Girardi et al. 2011, in prep).  That difference means that the actual recoverability of long-period binaries of all types would be somewhat lower than the bottom panel in Figure \ref{fig:aov1}.

\subsection{Accuracy of Eclipsing Binary Parameter Determination}

Neural networks will perform exceptionally well for light curves whose parameters are within the boundaries of the training set (especially in terms of non-linear interpolation), but the performance will deteriorate exponentially when extrapolation is required \citep{freeman1991}. To assure optimal operation, we trained the network on as wide a range of physically plausible light curves as possible (viz.~Fig.~\ref{fig:dists}).

The EBs with successfully recovered periods were passed to the artificial intelligence-based engine {\tt ebai} (Eclipsing Binaries via Artificial Intelligence; \citealt{prsa2008}) to derive physical parameters of the systems: $T_2/T_1$, $\rho_1+\rho_2$, $e \sin \omega$, $e \cos \omega$, and $\sin i$.  We first use {\tt polyfit} \citep{prsa2008} to find an analytical approximation for each light curve. This is necessary because the neural network takes flux values in equidistant phase steps as input. The {\tt polyfit} algorithm divides the phased light curve into $N$ intervals and fits an $n$-th order polynomial to the data in each interval; it then connects the polynomials into a chain by imposing two conditions: connectedness at chain knots and smooth chain wrapping on phase space boundaries. To find the best fit, {\tt polyfit} perturbs the position of chain knots; any knot positioning that improves the value of $\chi^2$ is adopted. This way a smooth polynomial chain with $N$ discontinuities in the first derivative represents EB data optimally. For the purpose of this simulation, second order polynomials were fit to 4 segments (i.e.~2 segments in eclipse and 2 segments between eclipses). The polyfits were then passed to {\tt ebai}.

Backpropagation network training, the only computationally intensive part of {\tt ebai}, needs to be performed only once for a given passband; this was done for the Sloan $r$ band on a 24-node Beowulf cluster using a parallelized version of the code -- the task took $\sim$2 days for 1 million iterations. Once trained, the network is able to process thousands of light curves in a fraction of a second on most computers. In particular, 10,000 light curves used in this simulation were processed in 0.5\,s on a 2.0GHz laptop, where most of this time was spent on I/O operations.

The initial test on a set of light curves with no noise or period uncertainty returned parameters with less than 10\% error for 75\% of all stars \citepalias[\S6.10, pg.~176]{LSSTbook}. The limiting factor is the limited number of visits that decreases the recoverability rate for long period EBs (due to a small number of data points in eclipses; cf.~Fig.~\ref{fig:lightcurves}).

\begin{figure*}[ht!]
\begin{center}
\includegraphics[height=0.48\textwidth, angle=270]{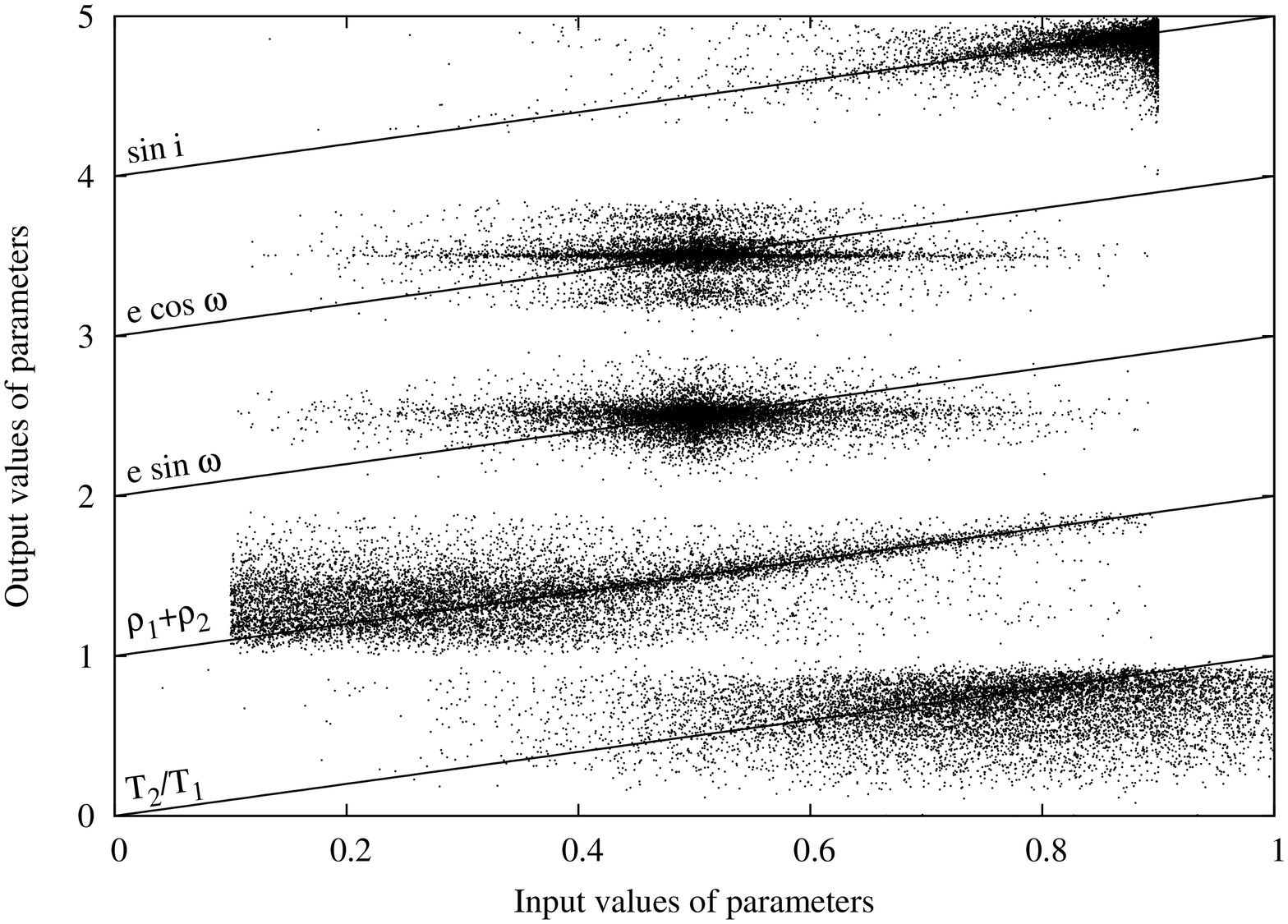}
\includegraphics[height=0.48\textwidth, angle=270]{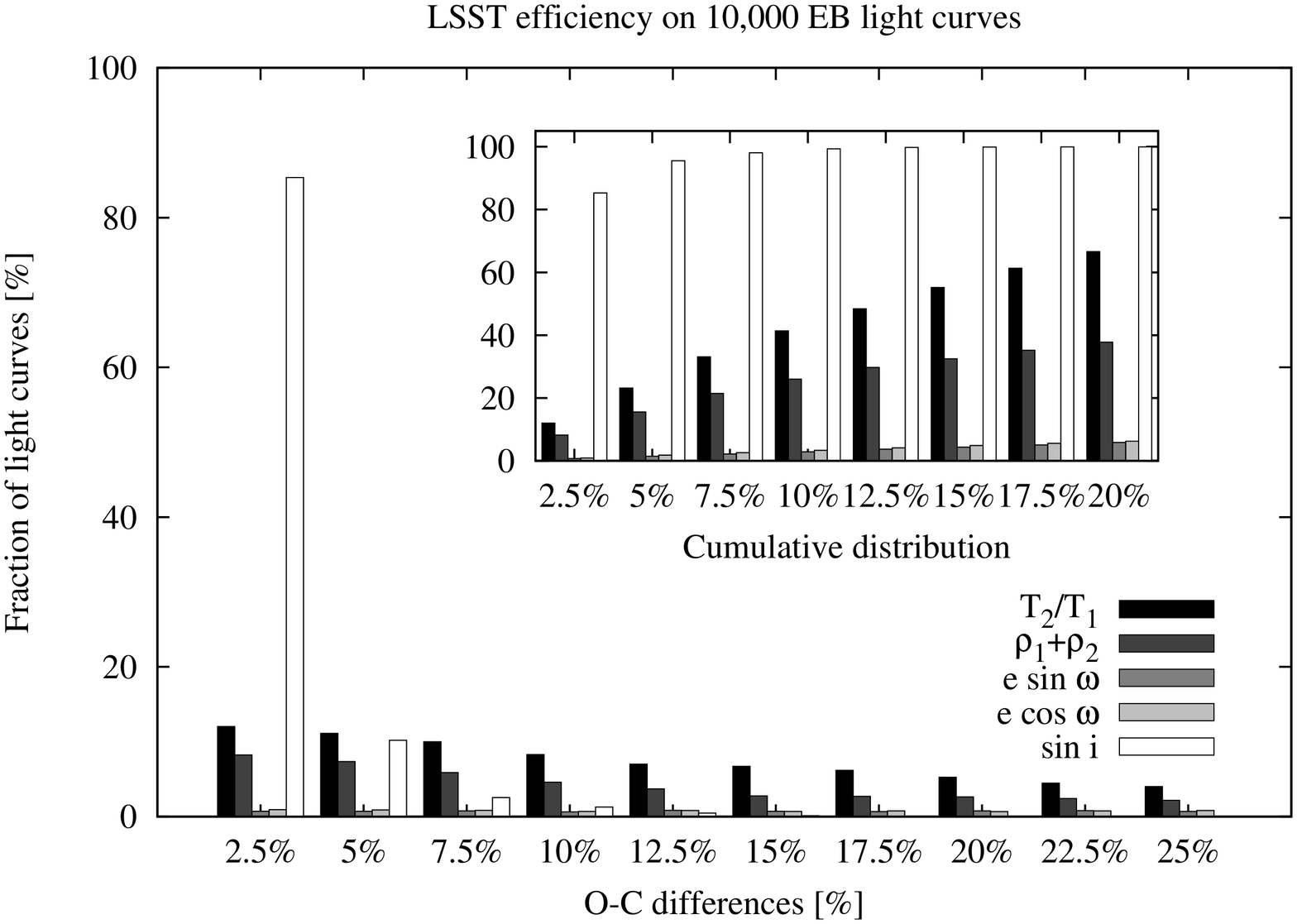} \\
\end{center}
\caption{{\bf Left:} comparison between input and {\tt ebai}-computed values of parameters. For neural network optimization purposes, parameter values were rescaled to the $[0.1, 0.9]$ interval. Successive parameters are vertically offset by 1.0 and correlation guidelines are provided to facilitate comparison. {\bf Right:} histogram of the percent residuals computed by {\tt ebai}. The inset depicts a cumulative distribution of the residuals. The recovery rate is $\sim$10\% to absolute errors better than 2.5\%, and $\sim$28\% to absolute errors better than 10\%. In comparison, a typical recovery rate for OGLE stars is $\sim$80\% to a 10\% precision \citep{prsa2008}, and $\sim$90\% to a 10\% precision for {\it Kepler} \citep{prsa2011}. The dominant reason for the lower recoverability rates for LSST when compared to OGLE and Kepler is a significantly smaller number of field visits and thus a poorer phase coverage -- cf.~Fig.~\ref{EB_fig_skymap}. \label{fig:ebai_results}}
\end{figure*}

The situation with added noise and period uncertainty becomes significantly worse. Fig.~\ref{fig:ebai_results} depicts the results of {\tt ebai} for the 10,000 simulated light curves used in this paper: $\sim$10\% of all parameters deviate from their true values by less than 2.5\%; $\sim$10\% by less than 5\%, and $\sim$28\% by less than 10\%. A 10\% error might seem large (typical error estimates of state-of-the-art EB modeling are close to 1-2\%), however {\tt ebai} serves to provide an \emph{initial} estimate for parameter values that would subsequently be improved by model-based methods such as Differential Corrections or Nelder \& Mead's Simplex, as implemented in PHOEBE. The simulation indicates that this will be readily possible for about a quarter of the sample.

\subsection{Overall Recovery} \label{sec:detrate}

\begin{figure}
\begin{center}
\includegraphics[height=\columnwidth, angle=0]{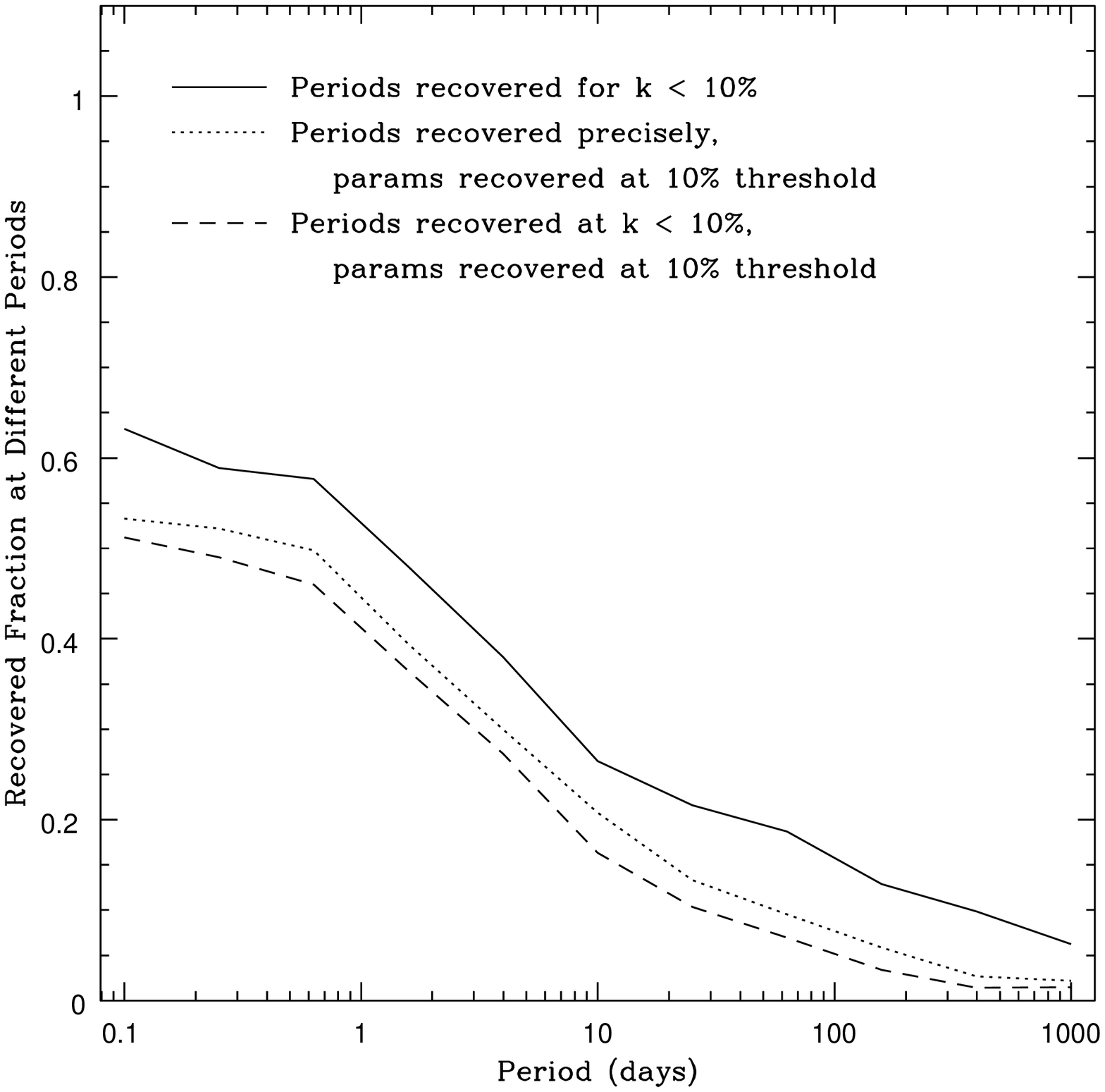} \\
\end{center}
\caption{The overall yield of eclipsing binary stars for LSST.  Periods are successfully recovered when the most discrepant points in the lightcurve are off by less than 10\% in phase ($k<10\%$, see \S \ref{sec:per_recover}). For parameter recovery the average of the percent difference between the true and recovered values for each of the five EB parameters is below 10\%. The solid curve is the same as in the bottom panel of Fig.~\ref{fig:aov1} and is reproduced here for convenience. It is evident that the bulk of suppressed recoverability comes from the failed period estimations. \label{fig:aov2}}
\end{figure}

Combining the AoV period recovery and the parameter recovery provides the overall LSST yield.  To set a threshold for ``correct'' parameter recovery, we compare the recovered value versus the true value for all 5 lightcurve parameters, and compute the percentage differences for each parameter.  We consider those systems for which the average of the five parameter percentage differences is below 10\% to be recovered.  The results are shown in Figure \ref{fig:aov2}.  As stated above in \ref{sec:per_recover}, the true recoverability fraction at longer periods would be lower than shown in \ref{fig:aov2}, due to the smaller proportion of giants compared to the fraction in our test lightcurves. Recoverability decrease towards longer periods is predominantly due to incorrectly determined periods; the suppression in recoverability due to failed parameter estimation does not significantly depend on the period of the binary.

\section{Conclusion}

The Large Synoptic Survey Telescope has the potential to dramatically impact the current state of stellar astrophysics. As the simulations described above demonstrate, LSST will detect a majority of the EBs with orbital periods less than 1 day, and substantial fractions of those with periods up to $\sim$100 days. This completeness estimate is based on analysis of a single passband; simultaneous analysis of all six LSST bands will in reality improve this completeness. With a nominal detection limit of $r=24.5$, a magnitude of $r=22.0$ should allow detection of targets with a $S/N$ of 10, $r=19.5$ will have $S/N$ of 100. Table \ref{EB_table_interesting_targets} shows the distance out to which certain fiducial EB types can be detected. For example, a pair of eclipsing M2 dwarfs will be detected out to 1 kpc with $S/N$ of 100. 

\begin{center}
\begin{table}[t]
\caption{Distance Limits for LSST Detection of Sample EBs.}
\scriptsize
\begin{tabular}{c|ccc}
Sample Binary${}^a$ Type & Binary Absolute Magnitude & Distance${}^b$ for $r=22.0$ [kpc] & Distance${}^b$ for $r=19.5$ [kpc] \\
\hline
M5V + M5V    & 12.9 &    0.7 &   0.2 \\
M2V + M2V    &  9.0 &    4.0 &   1.3 \\
K0V + K0V    &  5.0 &   25.1 &   7.9 \\
G2V + MxV    &  4.6 &   30.2 &   9.5 \\
G5III + GxV  &  2.9 &   66.1 &  20.1 \\
\hline
\multicolumn{4}{p{0.9\textwidth}}{${}^a$Scientifically interesting EB systems. EBs with M-dwarf components are rare in the literature. Their discovery will permit detailed testing of stellar models in this important mass regime. G-dwarf/M-dwarf pairs will be particularly valuable for pinning down the properties of M-dwarfs, since the temperature scale of G-dwarfs is relatively well established.  A particularly exciting prospect are Cepheids (G giants) in EB systems.} \\
\multicolumn{4}{p{0.9\textwidth}}{${}^b$Assuming no extinction.}
\end{tabular}
\label{EB_table_interesting_targets}
\end{table}
\end{center}

We can estimate the number of EBs that LSST will be able to fully characterize.  Experience modeling EB light curves shows that $S/N \sim 10$ per data point typically suffices for the determination of physical and geometric parameters to a few percent.  Gaia will observe $\sim$1 billion stars down to $r\sim20.5$ over the whole sky \citep{lindegren2008}.  We can expect that LSST will observe $\sim$0.5 billion stars to this same depth in the southern hemisphere; scaling this to $r \sim 22.0$ yields $\sim$2 billion stars.  Extrapolating the results from {\it Kepler} \citep{prsa2011} that observed 1879 EBs in the sample of 156,000 stars (1.2\%), the LSST sample will contain $\sim$24 million EBs with $S/N \geq 10$. The average detection rate (see section \ref{sec:detrate}) for EBs over all periods will be around 28\%, bringing the total number to $\sim$6.7 million EBs.  Roughly 25\% of those will have components of similar luminosities (double-lined systems), yielding $\sim$1.7 million EBs with $S/N \geq 10$ for ready detailed modeling.

The 28\% success rate seems low, but the following needs to be taken into consideration: 1) the recovery rate is better than 50\% for short period systems ($P < 1$-d) and can be improved to better than 60\% by optimizing the period finder; 2) the cases where the period finder and/or the neural network fail may be identified by simply evaluating the $\chi^2$ for each light curve, so that statistical results can be properly weighted. Thus, $\sim$7 million EBs with parameter uncertainties less than 10\% may be readily identified; and 3) while for the longer period EBs the inherent cadence limitation prevents recoverability, shorter period EBs will be passed to a physical model for differential corrections. These can improve the recoverability substantially whenever the initial {\tt ebai} parameter estimates are close (within 10\%) of the optimal solution.

While LSST will revolutionize the numbers of detected and characterized EBs, our simulations predict a strong selection effect w.r.t.~the orbital period. With only several hundred data points in the course of 10 year mission span, the probability of sampling a long period binary during eclipses becomes notably small and suppresses detection, classification and characterization success to under 10\%. Another identified deficiency of the sample due to a limited cadence is the lack of ability to characterize any intrinsic variability such as spots, accretion disks and other transient phenomena. Thus, we foresee that the main LSST legacy will be to make a census and statistical analysis of millions of galactic and extragalactic EBs, but the burden of careful object-by-object studies based on the prime astrophysical importance will fall on the follow-up program.

\begin{acknowledgements}

A.~Pr\v sa gratefully acknowledges support of the NSF/RUI Grant No. AST-05-07542. Computations are performed on a dedicated 24-node Beowulf cluster financed in part by the Astronomy Department of Villanova University. This work, and Vanderbilt University's participation in the LSST project, is made possible through the Vanderbilt Initiative in Data-intensive Astrophysics (VIDA). Funding for the SDSS and SDSS-II has been provided by the Alfred P. Sloan Foundation, the Participating Institutions, the National Science Foundation, the U.S. Department of Energy, the National Aeronautics and Space Administration, the Japanese Monbukagakusho, the Max Planck Society, and the Higher Education Funding Council for England. The SDSS Web Site is http://www.sdss.org/.

\end{acknowledgements}

\bibliographystyle{aa}
\bibliography{paper}

\begin{thebibliography}{27}
\expandafter\ifx\csname natexlab\endcsname\relax\def\natexlab#1{#1}\fi

\bibitem[{{Abazajian} {et~al.}(2009){Abazajian}, {Adelman-McCarthy},
  {Ag{\"{u}}eros}, {Allam}, {Prieto}, {An}, {Anderson}, {Anderson}, {Annis},
  {Bahcall}, {Bailer-Jones}, {Barentine}, {Bassett}, {Becker}, {Beers}, {Bell},
  {Belokurov}, {Berlind}, {Berman}, {Bernardi}, {Bickerton}, {Bizyaev},
  {Blakeslee}, {Blanton}, {Bochanski}, {Boroski}, {Brewington}, {Brinchmann},
  {Brinkmann}, {Brunner}, {Budav{\'{a}}ri}, {Carey}, {Carliles}, {Carr},
  {Castander}, {Cinabro}, {Connolly}, {Csabai}, {Cunha}, {Czarapata},
  {Davenport}, {Haas}, {Dilday}, {Doi}, {Eisenstein}, {Evans}, {Evans}, {Fan},
  {Friedman}, {Frieman}, {Fukugita}, {G{\"{a}}nsicke}, {Gates}, {Gillespie},
  {Gilmore}, {Gonzalez}, {Gonzalez}, {Grebel}, {Gunn}, {Gy{\"{o}}ry}, {Hall},
  {Harding}, {Harris}, {Harvanek}, {Hawley}, {Hayes}, {Heckman}, {Hendry},
  {Hennessy}, {Hindsley}, {Hoblitt}, {Hogan}, {Hogg}, {Holtzman}, {Hyde},
  {Ichikawa}, {Ichikawa}, {Im}, {Ivezi{\'{c}}}, {Jester}, {Jiang}, {Johnson},
  {Jorgensen}, {Juri{\'{c}}}, {Kent}, {Kessler}, {Kleinman}, {Knapp},
  {Konishi}, {Kron}, {Krzesinski}, {Kuropatkin}, {Lampeitl}, {Lebedeva}, {Lee},
  {Lee}, {Leger}, {L{\'{e}}pine}, {Li}, {Lima}, {Lin}, {Long}, {Loomis},
  {Loveday}, {Lupton}, {Magnier}, {Malanushenko}, {Malanushenko}, {Mandelbaum},
  {Margon}, {Marriner}, {Mart{'{\i}}nez-Delgado}, {Matsubara}, {McGehee},
  {McKay}, {Meiksin}, {Morrison}, {Mullally}, {Munn}, {Murphy}, {Nash},
  {Nebot}, {Neilsen}, {Newberg}, {Newman}, {Nichol}, {Nicinski},
  {Nieto-Santisteban}, {Nitta}, {Okamura}, {Oravetz}, {Ostriker}, {Owen},
  {Padmanabhan}, {Pan}, {Park}, {Pauls}, {Peoples}, {Percival}, {Pier}, {Pope},
  {Pourbaix}, {Price}, {Purger}, {Quinn}, {Raddick}, {Fiorentin}, {Richards},
  {Richmond}, {Riess}, {Rix}, {Rockosi}, {Sako}, {Schlegel}, {Schneider},
  {Scholz}, {Schreiber}, {Schwope}, {Seljak}, {Sesar}, {Sheldon}, {Shimasaku},
  {Sibley}, {Simmons}, {Sivarani}, {Smith}, {Smith}, {{C}i{\'{c}}}, {Snedden},
  {Stebbins}, {Steinmetz}, {Stoughton}, {Strauss}, {Rao}, {Suto}, {Szalay},
  {Szapudi}, {Szkody}, {Tanaka}, {Tegmark}, {Teodoro}, {Thakar}, {Tremonti},
  {Tucker}, {Uomoto}, {Berk}, {Vandenberg}, {Vidrih}, {Vogeley}, {Voges},
  {Vogt}, {Wadadekar}, {Watters}, {Weinberg}, {West}, {White}, {Wilhite},
  {Wonders}, {Yanny}, {Yocum}, {York}, {Zehavi}, {Zibetti}, \&
  {Zucker}}]{sdssdr7}
{Abazajian}, K.~N., {Adelman-McCarthy}, J.~K., {Ag{\"{u}}eros}, M.~A., {et~al.}
  2009, ApJ. Supp., 182, 543

\bibitem[{{Borucki} {et~al.}(2010){Borucki}, {Koch}, {Basri}, {Batalha},
  {Brown}, {Caldwell}, {Caldwell}, {Christensen-Dalsgaard}, {Cochran},
  {DeVore}, {Dunham}, {Dupree}, {Gautier}, {Geary}, {Gilliland}, {Gould},
  {Howell}, {Jenkins}, {Kondo}, {Latham}, {Marcy}, {Meibom}, {Kjeldsen},
  {Lissauer}, {Monet}, {Morrison}, {Sasselov}, {Tarter}, {Boss}, {Brownlee},
  {Owen}, {Buzasi}, {Charbonneau}, {Doyle}, {Fortney}, {Ford}, {Holman},
  {Seager}, {Steffen}, {Welsh}, {Rowe}, {Anderson}, {Buchhave}, {Ciardi},
  {Walkowicz}, {Sherry}, {Horch}, {Isaacson}, {Everett}, {Fischer}, {Torres},
  {Johnson}, {Endl}, {MacQueen}, {Bryson}, {Dotson}, {Haas}, {Kolodziejczak},
  {Van Cleve}, {Chandrasekaran}, {Twicken}, {Quintana}, {Clarke}, {Allen},
  {Li}, {Wu}, {Tenenbaum}, {Verner}, {Bruhweiler}, {Barnes}, \&
  {Prsa}}]{borucki2010}
{Borucki}, W.~J., {Koch}, D., {Basri}, G., {et~al.} 2010, Science, 327, 977

\bibitem[{{Cook} {et~al.}(2009){Cook}, {Pinto}, {Delgado}, {Miller}, {Petry},
  {Saha}, {Gee}, {Tyson}, {Ivezic}, {Jones}, \& {LSST
  Collaboration}}]{cook2009}
{Cook}, K.~H., {Pinto}, P.~A., {Delgado}, F., {et~al.} 2009, in Bulletin of the
  American Astronomical Society, Vol.~41, 367--+

\bibitem[{{Debosscher} {et~al.}(2007){Debosscher}, {Sarro}, {Aerts}, {Cuypers},
  {Vandenbussche}, {Garrido}, \& {Solano}}]{debosscher2007}
{Debosscher}, J., {Sarro}, L.~M., {Aerts}, C., {et~al.} 2007, \aap, 475, 1159

\bibitem[{{Devor}(2005)}]{devor2005}
{Devor}, J. 2005, ApJ, 628, 411

\bibitem[{Freeman \& Skapura(1991)}]{freeman1991}
Freeman, J. \& Skapura, D. 1991, {Neural networks: algorithms, applications,
  and programming techniques}, Computation and neural systems series
  (Addison-Wesley)

\bibitem[{{Guinan} {et~al.}(1998){Guinan}, {Fitzpatrick}, {Dewarf}, {Maloney},
  {Maurone}, {Ribas}, {Pritchard}, {Bradstreet}, \& {Gim{\'e}nez}}]{guinan1998}
{Guinan}, E.~F., {Fitzpatrick}, E.~L., {Dewarf}, L.~E., {et~al.} 1998, \apjl,
  509, L21

\bibitem[{Hartman {et~al.}(2009)Hartman, Bakos, Noyes, Sip\"{o}cz, Kov\'{a}cs,
  Mazeh, Shporer, \& P\'{a}l}]{hartman2009}
Hartman, J.~D., Bakos, Noyes, R.~W., {et~al.} 2009, ArXiv e-prints

\bibitem[{Hartman {et~al.}(2008)Hartman, Gaudi, Holman, McLeod, Stanek,
  Barranco, Pinsonneault, \& Kalirai}]{hartman2008}
Hartman, J.~D., Gaudi, B.~S., Holman, M.~J., {et~al.} 2008, ApJ, 675, 1254

\bibitem[{{Ivezic} {et~al.}(2008){Ivezic}, {Tyson}, {Allsman}, {Andrew},
  {Angel}, {Axelrod}, {Barr}, {Becker}, {Becla}, {Beldica}, {Blandford},
  {Brandt}, {Bullock}, {Burke}, {Chandrasekharan}, {Chesley}, {Claver},
  {Connolly}, {Cook}, {Cooray}, {Cribbs}, {Cutri}, {Daues}, {Delgado},
  {Ferguson}, {Geary}, {Gee}, {Gilmore}, {Gressler}, {Hogan}, {Huffer},
  {Jacoby}, {Jain}, {Jernigan}, {Jones}, {Juric}, {Kahn}, {Kalirai}, {Kantor},
  {Kirkby}, {Knox}, {Krabbendam}, {Krughoff}, {Kulkarni}, {Lambert}, {Levine},
  {Liang}, {Lim}, {Lupton}, {Marshall}, {Marshall}, {May}, {Miller}, {Mills},
  {Monet}, {Neill}, {Nordby}, {O'Connor}, {Oliver}, {Olivier}, {Owen},
  {Peterson}, {Petry}, {Pierfederici}, {Pietrowicz}, {Pike}, {Pinto}, {Plante},
  {Radeka}, {Rasmussen}, {Rosing}, {Saha}, {Schalk}, {Schindler}, {Schneider},
  {Schumacher}, {Sebag}, {Seppala}, {Shipsey}, {Silvestri}, {Smith}, {Smith},
  {Strauss}, {Stubbs}, {Sweeney}, {Szalay}, {Thaler}, {VandenBerk}, {Warner},
  {Willman}, {Wittman}, {Wolff}, {Wood-Vasey}, \& {Zhan}}]{ivezic2008}
{Ivezic}, Z., {Tyson}, J.~A., {Allsman}, R., {et~al.} 2008, ArXiv e-prints

\bibitem[{{Jedicke} \& {Pan-STARRS}(2007)}]{jedicke2007}
{Jedicke}, R. \& {Pan-STARRS}. 2007, in Bulletin of the American Astronomical
  Society, Vol.~38, 421--+

\bibitem[{{Jordi} {et~al.}(2010){Jordi}, {Gebran}, {Carrasco}, {de Bruijne},
  {Voss}, {Fabricius}, {Knude}, {Vallenari}, {Kohley}, \& {Mora}}]{jordi2010}
{Jordi}, C., {Gebran}, M., {Carrasco}, J.~M., {et~al.} 2010, \aap, 523, A48+

\bibitem[{{Kaiser}(2005)}]{kaiser2005}
{Kaiser}, N. 2005, in Probing the Dark Universe with Subaru and Gemini

\bibitem[{{Kaiser} {et~al.}(2002){Kaiser}, {Aussel}, {Burke}, {Boesgaard},
  {Chambers}, {Chun}, {Heasley}, {Hodapp}, {Hunt}, {Jedicke}, {Jewitt},
  {Kudritzki}, {Luppino}, {Maberry}, {Magnier}, {Monet}, {Onaka}, {Pickles},
  {Rhoads}, {Simon}, {Szalay}, {Szapudi}, {Tholen}, {Tonry}, {Waterson}, \&
  {Wick}}]{kaiser2002}
{Kaiser}, N., {Aussel}, H., {Burke}, B.~E., {et~al.} 2002, in Society of
  Photo-Optical Instrumentation Engineers (SPIE) Conference Series, ed. J.~A.
  {Tyson} \& S.~{Wolff}, Vol. 4836, 154--164

\bibitem[{Kov\'{a}cs {et~al.}(2002)Kov\'{a}cs, Zucker, \& Mazeh}]{Kovacs2002}
Kov\'{a}cs, G., Zucker, S., \& Mazeh, T. 2002, A\&A, 391, 369

\bibitem[{{Lindegren} {et~al.}(2008){Lindegren}, {Babusiaux}, {Bailer-Jones},
  {Bastian}, {Brown}, {Cropper}, {H{\o}g}, {Jordi}, {Katz}, {van Leeuwen},
  {Luri}, {Mignard}, {de Bruijne}, \& {Prusti}}]{lindegren2008}
{Lindegren}, L., {Babusiaux}, C., {Bailer-Jones}, C., {et~al.} 2008, in IAU
  Symposium, ed. W.~J. {Jin}, I.~{Platais}, \& M.~A.~C. {Perryman}, Vol. 248,
  217--223

\bibitem[{{LSST Science Collaborations} {et~al.}(2009){LSST Science
  Collaborations}, Abell, Allison, Anderson, Andrew, Angel, Armus, Arnett,
  Asztalos, Axelrod, Bailey, Ballantyne, Bankert, Barkhouse, Barr, Barrientos,
  Barth, Bartlett, Becker, Becla, Beers, Bernstein, Biswas, Blanton, Bloom,
  Bochanski, Boeshaar, Borne, Bradac, Brandt, Bridge, Brown, Brunner, Bullock,
  Burgasser, Burge, Burke, Cargile, Chandrasekharan, Chartas, Chesley, {Y.-H.
  Chu}, Cinabro, Claire, Claver, Clowe, Connolly, Cook, Cooke, Cooray, Covey,
  Culliton, de~Jong, de~Vries, Debattista, Delgado, Dell'Antonio, Dhital,
  Di~Stefano, Dickinson, Dilday, Djorgovski, Dobler, Donalek, Dubois-Felsmann,
  Durech, Eliasdottir, Eracleous, Eyer, Falco, Fan, Fassnacht, Ferguson,
  Fernandez, Fields, Finkbeiner, Figueroa, Fox, Francke, Frank, Frieman,
  Fromenteau, Furqan, Galaz, Gal-Yam, Garnavich, Gawiser, Geary, Gee, Gibson,
  Gilmore, Grace, Green, Gressler, Grillmair, Habib, Haggerty, Hamuy, Harris,
  Hawley, Heavens, Hebb, Henry, Hileman, Hilton, Hoadley, Holberg, Holman,
  Howell, Infante, Ivezic, Jacoby, Jain, {R}, {Jedicke}, Jee, Jernigan, Jha,
  Johnston, Jones, Juric, Kaasalainen, {Styliani}, {Kafka}, Kahn, Kaib,
  Kalirai, Kantor, Kasliwal, Keeton, Kessler, Knezevic, Kowalski, Krabbendam,
  Krughoff, Kulkarni, Kuhlman, Lacy, Lepine, Liang, Lien, Lira, Long, Lorenz,
  Lotz, Lupton, Lutz, Macri, Mahabal, Mandelbaum, Marshall, May, McGehee,
  Meadows, Meert, Milani, Miller, Miller, Mills, Minniti, Monet, Mukadam,
  Nakar, Neill, Newman, Nikolaev, Nordby, O'Connor, Oguri, Oliver, Olivier,
  Olsen, Olsen, Olszewski, Oluseyi, Padilla, Parker, Pepper, Peterson, Petry,
  Pinto, Pizagno, Popescu, Prsa, Radcka, Raddick, Rasmussen, Rau, Rho, Rhoads,
  Richards, Ridgway, Robertson, Roskar, Saha, Sarajedini, Scannapieco, Schalk,
  Schindler, Schmidt, Schmidt, Schneider, Schumacher, Scranton, Sebag, Seppala,
  Shemmer, Simon, Sivertz, Smith, Smith, Smith, Spitz, Stanford, Stassun,
  Strader, Strauss, Stubbs, Sweeney, Szalay, Szkody, Takada, Thorman, Trilling,
  Trimble, Tyson, Van~Berg, Berk, VanderPlas, Verde, Vrsnak, Walkowicz,
  Wandelt, Wang, Wang, Warner, Wechsler, West, Wiecha, Williams, Willman,
  Wittman, Wolff, Wood-Vasey, Wozniak, Young, Zentner, \& Zhan}]{LSSTbook}
{LSST Science Collaborations}, Abell, P.~A., Allison, J., {et~al.} 2009, ArXiv
  e-prints

\bibitem[{{Munari} {et~al.}(2001){Munari}, {Tomov}, {Zwitter}, {Milone},
  {Kallrath}, {Marrese}, {Boschi}, {Pr{\v s}a}, {Tomasella}, \&
  {Moro}}]{munari2001}
{Munari}, U., {Tomov}, T., {Zwitter}, T., {et~al.} 2001, \aap, 378, 477

\bibitem[{{Pr{\v s}a} {et~al.}(2011){Pr{\v s}a}, {Batalha}, {Slawson}, {Doyle},
  {Welsh}, {Orosz}, {Seager}, {Rucker}, {Mjaseth}, {Engle}, {Conroy},
  {Jenkins}, {Caldwell}, {Koch}, \& {Borucki}}]{prsa2011}
{Pr{\v s}a}, A., {Batalha}, N., {Slawson}, R.~W., {et~al.} 2011, \aj, 141, 83

\bibitem[{{Pr{\v s}a} {et~al.}(2008){Pr{\v s}a}, {Guinan}, {Devinney},
  {DeGeorge}, {Bradstreet}, {Giammarco}, {Alcock}, \& {Engle}}]{prsa2008}
{Pr{\v s}a}, A., {Guinan}, E.~F., {Devinney}, E.~J., {et~al.} 2008, \apj, 687,
  542

\bibitem[{{Pr{\v s}a} \& {Zwitter}(2005)}]{prsa2005}
{Pr{\v s}a}, A. \& {Zwitter}, T. 2005, \apj, 628, 426

\bibitem[{Schwarzenberg-Czerny(1989)}]{s-c89}
Schwarzenberg-Czerny, A. 1989, MNRAS, 241, 153

\bibitem[{{Stassun} {et~al.}(2009){Stassun}, {Hebb}, {L{\'o}pez-Morales}, \&
  {Pr{\v s}a}}]{stassun2009}
{Stassun}, K.~G., {Hebb}, L., {L{\'o}pez-Morales}, M., \& {Pr{\v s}a}, A. 2009,
  in IAU Symposium, ed. {E.~E.~Mamajek, D.~R.~Soderblom, \& R.~F.~G.~Wyse},
  Vol. 258, 161--170

\bibitem[{{Thompson} {et~al.}(2010){Thompson}, {Kaluzny}, {Rucinski},
  {Krzeminski}, {Pych}, {Dotter}, \& {Burley}}]{thompson2010}
{Thompson}, I.~B., {Kaluzny}, J., {Rucinski}, S.~M., {et~al.} 2010, \aj, 139,
  329

\bibitem[{{Torres} {et~al.}(2010){Torres}, {Andersen}, \&
  {Gim{\'{e}}nez}}]{torres2010}
{Torres}, G., {Andersen}, J., \& {Gim{\'{e}}nez}, A. 2010, Astron. Astrophys.
  Rev., 18, 67

\bibitem[{{van Kerkwijk} {et~al.}(2010){van Kerkwijk}, {Rappaport}, {Breton},
  {Justham}, {Podsiadlowski}, \& {Han}}]{vankerkwijk2010}
{van Kerkwijk}, M.~H., {Rappaport}, S.~A., {Breton}, R.~P., {et~al.} 2010,
  \apj, 715, 51

\bibitem[{{Wilson} \& {Devinney}(1971)}]{wd1971}
{Wilson}, R.~E. \& {Devinney}, E.~J. 1971, \apj, 166, 605

\end{thebibliography}

\end{document}